\documentclass[conference]{IEEEtran}

\usepackage{cite}
\usepackage{amsmath,amssymb,amsfonts}
\usepackage{graphicx}
\usepackage{textcomp}
\usepackage{xcolor}
\usepackage{booktabs}
\usepackage{multirow}
\usepackage{hyperref}
\usepackage{url}
\usepackage{algorithm}
\usepackage{algpseudocode}
\usepackage{tabularx}
\usepackage[caption=false]{subfig}

\def\BibTeX{{\rm B\kern-.05em{\sc i\kern-.025em b}\kern-.08em
    T\kern-.1667em\lower.7ex\hbox{E}\kern-.125emX}}

\begin{document}

\title{Hybrid Retrieval for COVID-19 Literature:\\Comparing Rank Fusion and Projection Fusion with Diversity Reranking}

\author{
\IEEEauthorblockN{Harishkumar Kishorkumar Prajapati, Deniz Genco Atilla, Boran Sac}
\IEEEauthorblockA{
School of Electronic Engineering and Computer Science\\
Queen Mary University of London, London, UK\\
\texttt{p.harishkumarkishorkumar@se25.qmul.ac.uk, d.atilla@se25.qmul.ac.uk, b.sac@se25.qmul.ac.uk}}
}

\maketitle

\begin{abstract}
We present a hybrid retrieval system for COVID-19 scientific literature, evaluated on the TREC-COVID benchmark (171,332 papers, 50 expert queries). The system implements six retrieval configurations spanning sparse (SPLADE), dense (BGE), rank-level fusion (RRF), and a projection-based vector fusion (B5) approach. RRF fusion achieves the best relevance (nDCG@10\,=\,0.828), outperforming dense-only by 6.1\% and sparse-only by 14.9\%. Our projection fusion variant reaches nDCG@10\,=\,0.678 on expert queries while being 33\% faster (847\,ms vs.\ 1271\,ms) and producing 2.2$\times$ higher ILD@10 than RRF. Evaluation across 400 queries---including expert, machine-generated, and three paraphrase styles---shows that B5 delivers the largest relative gain on keyword-heavy reformulations (+8.8\%), although RRF remains best in absolute nDCG@10. On expert queries, MMR reranking increases intra-list diversity by 23.8--24.5\% at a 20.4--25.4\% nDCG@10 cost. Both fusion pipelines evaluated for latency remain below the sub-2\,s target across all query sets. The system is deployed as a Streamlit web application backed by Pinecone serverless indices.
\end{abstract}


\section{Introduction}

The COVID-19 pandemic produced an unprecedented volume of scientific publications, with the CORD-19 corpus exceeding 600,000 papers by mid-2022~\cite{wang2020cord19}. Clinicians and researchers face two interrelated challenges: (1)~keyword-only retrieval misses semantically related papers when medical terminology varies across studies, and (2)~top-ranked results frequently contain near-duplicate findings, reducing the diversity of information presented.

This paper describes a hybrid retrieval system built on the TREC-COVID benchmark~\cite{voorhees2020trec} that addresses both challenges. We implement and compare six retrieval configurations:

\begin{itemize}
    \item \textbf{B1} --- Sparse-only retrieval using SPLADE~\cite{formal2021splade}.
    \item \textbf{B2} --- Dense-only retrieval using BGE embeddings from the FlagEmbedding project~\cite{xiao2023bge}.
    \item \textbf{B4} --- Two-pass Reciprocal Rank Fusion (RRF)~\cite{cormack2009rrf} combining B1 and B2.
    \item \textbf{B3} --- RRF + Maximal Marginal Relevance (MMR) reranking~\cite{carbonell1998mmr}.
    \item \textbf{B5} --- Projection-based vector fusion using Achlioptas sparse random projection~\cite{achlioptas2003database}.
    \item \textbf{B5+MMR} --- B5 with MMR reranking.
\end{itemize}

We formulate three research questions:

\begin{itemize}
    \item[\textbf{RQ1}] Does hybrid fusion outperform single-method approaches?
    \item[\textbf{RQ2}] Does projection-based fusion offer competitive performance compared to RRF?
    \item[\textbf{RQ3}] How effective is MMR for reducing result redundancy?
\end{itemize}

\section{Related Work}

\subsection{Sparse and Dense Retrieval}
Traditional sparse retrieval relies on lexical matching through models such as BM25~\cite{robertson2009bm25}. SPLADE~\cite{formal2021splade, formal2022splade2} advances this paradigm by using a BERT-based masked language model to produce learned sparse representations over the full vocabulary, enabling implicit query expansion. Dense retrieval~\cite{karpukhin2020dpr} encodes queries and documents into fixed-dimensional vectors and retrieves by nearest-neighbour search. BGE (BAAI General Embeddings), released through the FlagEmbedding/C-Pack project~\cite{xiao2023bge}, is a recent contrastive-trained encoder producing 768-dimensional L2-normalised vectors.

\subsection{Hybrid Fusion Strategies}
Luan et al.~\cite{luan2021sparse} demonstrate that combining sparse and dense representations improves retrieval effectiveness. Reciprocal Rank Fusion (RRF)~\cite{cormack2009rrf} provides a simple, effective method for merging ranked lists without requiring score normalisation, using the formula $\text{RRF}(d) = \sum_{r \in R} \frac{w_r}{k + \text{rank}_r(d)}$ where $k=60$.

\subsection{Random Projection for Dimensionality Reduction}
The Johnson-Lindenstrauss lemma~\cite{johnson1984extensions} guarantees that random linear projections approximately preserve pairwise distances. Achlioptas~\cite{achlioptas2003database} showed that the projection matrix can be sparse (density $1/3$, entries $\pm\sqrt{3}$), reducing computational cost while maintaining this guarantee for the raw projection step. We apply such a projection to map SPLADE's 30,522-dimensional sparse vectors into BGE's 768-dimensional space before the subsequent normalisation and fusion stages, whose effectiveness we evaluate empirically.

\subsection{Diversity-Aware Reranking}
Maximal Marginal Relevance (MMR)~\cite{carbonell1998mmr} iteratively selects documents that are both relevant to the query and dissimilar to already-selected results, controlled by a parameter $\lambda$. Clarke et al.~\cite{clarke2008novelty} established diversity as a formal evaluation dimension in information retrieval.

\section{System Architecture}

\begin{figure}[htbp]
\centering
\IfFileExists{figures/architecture.png}{%
  \includegraphics[width=\columnwidth]{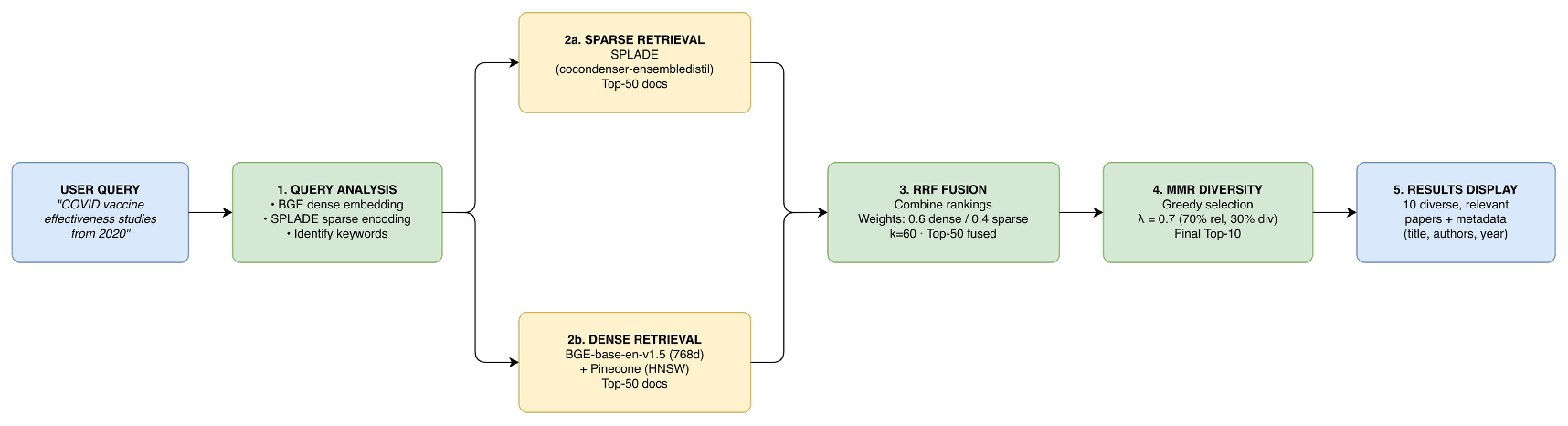}%
}{%
  \framebox[\columnwidth]{\rule{0pt}{4cm}\textit{Architecture diagram — see supplementary materials.}}%
}
\caption{RRF retrieval pipeline. A user query is encoded with both SPLADE (sparse) and BGE (dense) encoders. The two ranked lists of 50 candidates each are fused via Reciprocal Rank Fusion ($k{=}60$, weights $0.6/0.4$ for dense/sparse), then optionally reranked with MMR ($\lambda{=}0.7$) to produce 10 diverse, relevant results. The deployed system also includes a separate single-pass B5 path described in Section~\ref{sec:b5}.}
\label{fig:architecture}
\end{figure}

Figure~\ref{fig:architecture} illustrates the RRF retrieval pipeline, one of the two deployed retrieval paths. The companion B5 path uses the same dense and sparse encoders but projects SPLADE vectors into dense space, forms a single fused vector, and queries the separate \texttt{trec-covid-b5} index in one pass.

\subsection{Data Preparation}
The corpus comprises 171,332 papers from the TREC-COVID collection (CORD-19 snapshot, July 2020). Each document's title and abstract are concatenated into a single text field for encoding. Empty records are removed during preprocessing.

\subsection{Encoding Pipeline}

\subsubsection{Dense Encoding}
We use \texttt{BAAI/bge-base-en-v1.5} to encode documents and queries into 768-dimensional L2-normalised vectors. Queries are prefixed with \textit{``Represent this sentence for searching relevant passages:''} following BGE's recommended practice.

\subsubsection{Sparse Encoding}
We use \texttt{naver/splade-cocondenser-ensembledistil} to produce sparse vectors over the 30,522-token BERT vocabulary. The SPLADE aggregation applies $\log(1 + \text{ReLU}(\text{logits}))$ followed by max-pooling across token positions, yielding a sparse vector where each non-zero entry corresponds to an activated vocabulary term.

Due to the computational cost of SPLADE on 171K documents, all sparse vectors are pre-computed offline and cached as pickle files.

\subsection{Indexing}
We use Pinecone serverless with two indices:

\begin{itemize}
    \item \textbf{\texttt{trec-covid}}: Hybrid index storing both dense (768-d) and sparse vectors per document. Supports separate dense-only, sparse-only, and hybrid queries via a hybrid-index mixing parameter $\alpha_{\text{hyb}}$.
    \item \textbf{\texttt{trec-covid-b5}}: Dense-only index (768-d, dot product) storing pre-combined B5 projection vectors.
\end{itemize}

Indexing uses checkpointing to survive interruptions, processing documents in batches of 100. We initially used ChromaDB but migrated to Pinecone after exceeding the free-tier storage limit.

\subsection{Retrieval Methods}

\subsubsection{Sparse-Only (B1)}
Queries are encoded with SPLADE and sent to the hybrid index with $\alpha_{\text{hyb}}=0.0$, retrieving by sparse signal alone.

\subsubsection{Dense-Only (B2)}
Queries are encoded with BGE and sent with $\alpha_{\text{hyb}}=1.0$, retrieving by dense cosine similarity.

\subsubsection{RRF Fusion (B4)}
Two separate passes---one dense ($\alpha_{\text{hyb}}=1.0$), one sparse ($\alpha_{\text{hyb}}=0.0$)---each retrieve $K=50$ candidates. Results are fused using RRF with $k=60$ and weights $[0.6, 0.4]$ for dense and sparse respectively. The top-10 results are returned.

\subsubsection{RRF + MMR (B3)}
After RRF fusion, the top-50 candidates are reranked using MMR with $\lambda=0.7$ (70\% relevance, 30\% diversity). Document vectors are fetched from Pinecone for pairwise similarity computation.

\subsubsection{Projection Fusion (B5)}\label{sec:b5}
Our projection-fusion approach fuses sparse and dense representations at the vector level before retrieval. The pipeline is:

\begin{enumerate}
    \item Encode query with BGE $\rightarrow$ dense vector $\mathbf{d} \in \mathbb{R}^{768}$
    \item Encode query with SPLADE $\rightarrow$ sparse vector $\mathbf{s} \in \mathbb{R}^{30522}$
    \item Project: $\mathbf{p} = R \cdot \mathbf{s}$, where $R \in \mathbb{R}^{768 \times 30522}$ is the Achlioptas sparse random projection matrix (density $= 1/3$, non-zero entries $\pm\sqrt{3}$)
    \item L2-normalise $\mathbf{p}$
    \item Combine: $\mathbf{q} = \alpha_{\text{query}} \cdot \hat{\mathbf{d}} + (1-\alpha_{\text{query}}) \cdot \hat{\mathbf{p}}$, then L2-normalise
    \item Single dot-product query against the B5 index
\end{enumerate}

Document vectors are pre-combined analogously using $\alpha_{\text{doc}}$. We index documents at $\alpha_{\text{doc}} = 0.50$ (balanced). Query-side mixing is tuned independently without re-indexing; we select $\alpha_{\text{query}} = 0.95$ based on ablation (Section~\ref{sec:alpha}).

The projection matrix requires no training data and is theoretically motivated by Johnson-Lindenstrauss-style distance preservation at the raw projection stage. The subsequent normalisation and dense--projection fusion steps are evaluated empirically. The sparse matrix structure (2/3 zero entries) enables efficient multiplication.

\subsubsection{B5 + MMR}
Identical to B5 with MMR reranking ($\lambda=0.7$) applied to the top-50 single-pass candidates, using the combined B5 vector as the query reference for diversity computation.

\subsection{User Interface}
The system is deployed as a Streamlit web application with a sidebar for index selection (\texttt{trec-covid} or \texttt{trec-covid-b5}). Search results display titles, abstracts, relevance scores, and metadata.

\section{Experimental Setup}

\subsection{Query Sets}
We evaluate across five query sets totalling 400 queries:

\begin{itemize}
    \item \textbf{Expert 50}: Official TREC-COVID Round~5 topics with graded relevance judgments (0/1/2).
    \item \textbf{Generated 200}: Randomly sampled (seed=42) from BeIR/trec-covid-generated-queries (doc2query/T5). Binary relevance---each query maps to exactly one source document.
    \item \textbf{Paraphrase A (50)}: Conversational style---natural, everyday language with added context.
    \item \textbf{Paraphrase B (50)}: Semi-technical---domain-specific medical terminology and precise wording.
    \item \textbf{Paraphrase C (50)}: Keyword-heavy---concise, fewer words, keyword-focused phrasing.
\end{itemize}

Paraphrases A, B, and C were generated from the 50 expert queries using Claude Sonnet and inherit the original graded relevance judgments.

Evaluation uses a qrel set matched to each query source. Expert 50 uses the official TREC-COVID Round~5 graded qrels. Paraphrases A/B/C reuse those same graded qrels because each paraphrase is a reformulation of an expert topic. Generated 200 uses a separate binary qrel file built from the source-document mapping in the generated-query dataset, assigning one positive document to each query.

\subsection{Metrics}
\begin{itemize}
    \item \textbf{nDCG@10}: Primary metric---normalised discounted cumulative gain with graded relevance.
    \item \textbf{P@10}: Precision at 10.
    \item \textbf{MRR@10}: Mean reciprocal rank of the first relevant document.
    \item \textbf{MAP@10}: Mean average precision.
    \item \textbf{HitRate@10}: Fraction of queries with $\geq$1 relevant document in top-10.
    \item \textbf{ILD@10}: Intra-list diversity---mean pairwise cosine dissimilarity among top-10 results. Higher values indicate more diverse result sets.
\end{itemize}

All metrics are computed using \texttt{pytrec\_eval} against the corresponding qrel set for each query collection.

\subsection{Infrastructure}
Experiments run on Apple Silicon (MPS) with PyTorch. Pinecone serverless (free tier) hosts both indices. Latency is measured end-to-end from query encoding through result retrieval.

\section{Results}

\subsection{RQ1: Hybrid Fusion vs.\ Single Methods}

Table~\ref{tab:expert} presents results on the 50 expert queries. RRF fusion (B4) achieves the highest nDCG@10 of 0.8282, outperforming dense-only (B2) by +6.1\% and sparse-only (B1) by +14.9\%. Dense retrieval captures semantic similarity while SPLADE provides exact term matching---combining both signals consistently yields superior performance.

\begin{table}[htbp]
\caption{Expert 50 Queries --- All Systems}
\label{tab:expert}
\centering
\begin{tabular}{lcccc}
\toprule
System & nDCG@10 & P@10 & MRR@10 & ILD@10 \\
\midrule
B1 Sparse & 0.7203 & 0.7800 & 0.9117 & 0.1985 \\
B2 Dense & 0.7803 & 0.8280 & 0.9290 & 0.1781 \\
\textbf{B4 RRF} & \textbf{0.8282} & \textbf{0.8740} & \textbf{0.9800} & 0.1760 \\
B3 RRF+MMR & 0.6589 & 0.7000 & 0.9163 & 0.2191 \\
B5 Proj. & 0.6779 & 0.7180 & 0.8967 & 0.3890 \\
B5+MMR & 0.5056 & 0.5200 & 0.9033 & \textbf{0.4813} \\
\bottomrule
\end{tabular}
\end{table}

\subsection{RQ2: Projection Fusion vs.\ RRF}

B5 projection fusion achieves nDCG@10\,=\,0.6779 on expert queries---18.2\% below RRF. However, B5 offers three distinct advantages:

\begin{enumerate}
    \item \textbf{Speed}: 33\% faster on expert queries (847\,ms vs.\ 1271\,ms average latency) due to single-pass retrieval eliminating one Pinecone round-trip and the RRF fusion step.
    \item \textbf{Diversity}: 2.2$\times$ higher ILD@10 (0.389 vs.\ 0.176) without any explicit diversity mechanism. The projected vector space naturally produces more varied retrievals.
    \item \textbf{Keyword robustness}: B5 shows the largest relative gain on keyword-heavy paraphrases (+8.8\%), although RRF remains best in absolute nDCG@10 (0.8083) and both dense-only (+3.0\%) and sparse-only (+1.6\%) also improve (Section~\ref{sec:paraphrase}).
\end{enumerate}

\subsubsection{Alpha Tuning Ablation}\label{sec:alpha}

Table~\ref{tab:alpha} shows the effect of the query-side mixing parameter $\alpha_{\text{query}}$ on B5 performance (documents always indexed at $\alpha_{\text{doc}}=0.50$).

\begin{table}[htbp]
\caption{B5 Query-Side Alpha Tuning --- Expert 50 Queries}
\label{tab:alpha}
\centering
\begin{tabular}{cccc}
\toprule
$\alpha_{\text{query}}$ & nDCG@10 & P@10 & ILD@10 \\
\midrule
0.50 & 0.4752 & 0.5000 & 0.3481 \\
0.80 & 0.6265 & 0.6600 & 0.3705 \\
\textbf{0.95} & \textbf{0.6779} & \textbf{0.7180} & 0.3890 \\
1.00 & 0.6766 & 0.7160 & 0.3958 \\
\bottomrule
\end{tabular}
\end{table}

Higher $\alpha_{\text{query}}$ (more dense weight at query time) consistently improves relevance in this mixed-document setting. We select $\alpha_{\text{query}}=0.95$ as the best operating point: it slightly outperforms the $\alpha_{\text{query}}=1.0$ mixed-document setting by +0.19\% without requiring re-indexing. Because documents remain indexed at $\alpha_{\text{doc}}=0.50$, the $\alpha_{\text{query}}=1.0$ row is not a pure-dense baseline; establishing that comparison would require a separate $\alpha_{\text{doc}}=1.0$, $\alpha_{\text{query}}=1.0$ index. This asymmetric tuning---balanced documents, dense-biased queries---is therefore best understood as a practical operating choice rather than a proof against a true dense-only baseline.

\subsection{RQ3: MMR Diversity Trade-off}

Table~\ref{tab:mmr} quantifies the expert-query MMR trade-off. For both fusion pipelines, MMR increases diversity (ILD@10) by 23.8--24.5\% while reducing nDCG@10 by 20.4--25.4\%. Notably, B5 without MMR already achieves 2.2$\times$ higher diversity than RRF without MMR, showing that projection fusion is associated with more diverse result lists in this experimental setting.

\begin{table}[htbp]
\caption{MMR Diversity Trade-off --- Expert 50 Queries}
\label{tab:mmr}
\centering
\begin{tabular}{lcccc}
\toprule
System & nDCG@10 & ILD@10 & $\Delta$nDCG & $\Delta$ILD \\
\midrule
B4 RRF & 0.8282 & 0.1760 & --- & --- \\
B3 RRF+MMR & 0.6589 & 0.2191 & $-$20.4\% & +24.5\% \\
\midrule
B5 Proj. & 0.6779 & 0.3890 & --- & --- \\
B5+MMR & 0.5056 & 0.4813 & $-$25.4\% & +23.8\% \\
\bottomrule
\end{tabular}
\end{table}

\subsection{Generalisation: Generated Queries}

On 200 machine-generated queries (Table~\ref{tab:gen}), sparse-only SPLADE wins (nDCG@10\,=\,0.4272), reversing the expert-query pattern. Generated queries appear more keyword-like, which is consistent with stronger exact lexical matching on this set. Because this set uses binary source-document qrels rather than the graded TREC-COVID qrels used for Expert 50 and the paraphrases, these scores are best interpreted as a separate generalisation test rather than a directly comparable benchmark against the expert-query setting.

\begin{table}[htbp]
\caption{Generated 200 Queries}
\label{tab:gen}
\centering
\begin{tabular}{lcc}
\toprule
System & nDCG@10 & HitRate@10 \\
\midrule
\textbf{B1 Sparse} & \textbf{0.4272} & \textbf{0.5450} \\
B4 RRF & 0.3983 & 0.5300 \\
B2 Dense & 0.3829 & 0.5150 \\
B3 RRF+MMR & 0.3584 & 0.4600 \\
B5 Proj. & 0.3277 & 0.4450 \\
\bottomrule
\end{tabular}
\end{table}

\subsection{Paraphrase Robustness}\label{sec:paraphrase}

Table~\ref{tab:para} evaluates robustness under three query reformulation styles for the four non-MMR systems. Key findings:

\begin{itemize}
    \item \textbf{Conversational (A)}: SPLADE is most stable ($-$2.3\%). Looser natural-language phrasing affects dense and hybrid systems more.
    \item \textbf{Semi-technical (B)}: B5 is most stable ($-$8.4\%), and dense-only also remains comparatively stable ($-$9.9\%), while RRF ($-$13.6\%) and SPLADE ($-$14.2\%) degrade more sharply.
    \item \textbf{Keyword-heavy (C)}: B5 shows the largest relative gain (+8.8\%), but RRF remains best in absolute nDCG@10 (0.8083); dense-only (+3.0\%) and SPLADE (+1.6\%) also improve relative to the original wording.
\end{itemize}

\begin{table}[htbp]
\caption{Paraphrase Robustness --- nDCG@10 ($\Delta$ vs.\ Original)}
\label{tab:para}
\centering
\resizebox{\columnwidth}{!}{%
\begin{tabular}{lcccc}
\toprule
Style & B4 RRF & B5 Proj. & B2 Dense & B1 SPLADE \\
\midrule
Original & 0.8282 & 0.6779 & 0.7803 & 0.7203 \\
A Conv. & 0.7631 ($-$7.9\%) & 0.6413 ($-$5.4\%) & 0.6895 ($-$11.6\%) & 0.7039 ($-$2.3\%) \\
B Semi. & 0.7154 ($-$13.6\%) & 0.6206 ($-$8.4\%) & 0.7032 ($-$9.9\%) & 0.6179 ($-$14.2\%) \\
\textbf{C Key.} & \textbf{0.8083 ($-$2.4\%)} & 0.7373 (+8.8\%) & 0.8039 (+3.0\%) & 0.7315 (+1.6\%) \\
\bottomrule
\end{tabular}
}
\end{table}

\subsection{Latency Analysis}

Table~\ref{tab:latency} compares end-to-end latency for the two fusion pipelines. B5 is consistently faster because it requires only one Pinecone query vs.\ two for RRF. Both pipelines remain below the 2\,s target across all query sets.

\begin{table}[htbp]
\caption{Latency Comparison (ms) --- Avg / P95}
\label{tab:latency}
\centering
\begin{tabular}{lccc}
\toprule
Query Set & RRF (2-pass) & B5 (1-pass) & Speedup \\
\midrule
Expert 50 & 1271 / 1623 & 847 / 1111 & $-$33\% \\
Generated 200 & 844 / 1034 & 672 / 906 & $-$20\% \\
Para A & 969 / 1369 & 665 / 885 & $-$31\% \\
Para B & 914 / 1257 & 768 / 927 & $-$16\% \\
Para C & 988 / 1535 & 664 / 865 & $-$33\% \\
\midrule
\textbf{Average} & \textbf{997 / 1364} & \textbf{723 / 939} & \textbf{$-$27\%} \\
\bottomrule
\end{tabular}
\end{table}

\subsection{Consolidated Results}

Table~\ref{tab:consolidated} presents nDCG@10 across all five query sets.

\begin{table}[htbp]
\caption{nDCG@10 Across All Query Sets. The Avg column is descriptive only because Generated 200 uses different qrels.}
\label{tab:consolidated}
\centering
\begin{tabular}{lcccccc}
\toprule
System & Exp. & Gen. & A & B & C & Avg* \\
\midrule
B1 Sparse & .720 & \textbf{.427} & .704 & .618 & .732 & .640 \\
B2 Dense & .780 & .383 & .690 & .703 & .804 & .672 \\
\textbf{B4 RRF} & \textbf{.828} & .398 & \textbf{.763} & \textbf{.715} & \textbf{.808} & \textbf{.703} \\
B3 RRF+MMR & .659 & .358 & .623 & .616 & .668 & .585 \\
B5 Proj. & .678 & .328 & .641 & .621 & .737 & .601 \\
\bottomrule
\end{tabular}
\end{table}

Because Generated 200 uses binary source-document qrels whereas the other four sets use graded TREC-COVID qrels, the Avg* column should be interpreted as a descriptive summary rather than a directly comparable aggregate benchmark.

\section{Discussion}

\subsection{Fusion Strategy Trade-offs}
Table~\ref{tab:consolidated} shows a clear split by query type. On the four query sets scored with graded TREC-COVID qrels (Expert 50 and paraphrase sets A--C), RRF achieves the best absolute nDCG@10. This supports RQ1: combining sparse and dense signals is beneficial for expert-written and reformulated queries. By contrast, on Generated 200, sparse-only SPLADE performs best, suggesting that these machine-generated queries are more keyword-oriented and therefore benefit more from lexical matching. Because Generated 200 uses binary source-document qrels, we interpret it as a separate generalisation test rather than as a directly comparable benchmark against the graded-qrel settings.

\begin{itemize}
    \item \textbf{Effectiveness-first}: Use RRF when maximising relevance is the main objective; it leads Expert 50 (0.828) and paraphrase sets A/B/C (0.763/0.715/0.808).
    \item \textbf{Keyword-like inputs}: Use SPLADE when queries resemble terse keyword searches; it is the best system on Generated 200 (0.427).
    \item \textbf{Latency/diversity-sensitive settings}: Use B5 when a single-pass pipeline and broader result coverage matter more than peak nDCG@10. On Expert 50, B5 reduces latency from 1271\,ms to 847\,ms and raises ILD@10 from 0.176 to 0.389, while remaining relatively competitive on the keyword-heavy paraphrases (0.737 vs.\ 0.808 for RRF).
\end{itemize}

Taken together, these results answer RQ2 more cautiously than RQ1. B5 is not the top-scoring system in absolute nDCG@10 on any of the reported query sets, but it does offer a distinct operating point that trades some effectiveness for lower latency and substantially higher diversity.

\subsection{Projection Fusion Analysis}
The selected operating point uses balanced document vectors ($\alpha_{\text{doc}}=0.50$) and dense-biased query vectors ($\alpha_{\text{query}}=0.95$). Within the mixed-document setting of Table~\ref{tab:alpha}, $\alpha_{\text{query}}=0.95$ exceeds $\alpha_{\text{query}}=1.0$ by +0.19\%. We therefore interpret the ablation as evidence that a small projected component can still help within the chosen mixed-document index. It should not be read as evidence that projection fusion generally outperforms a separately indexed pure-dense baseline, because the document representation is held fixed throughout this comparison. A document-side ablation would be required to justify a stronger claim about the optimality of $\alpha_{\text{doc}}=0.50$.

\subsection{Observed Diversity Pattern}
B5 achieves ILD@10\,=\,0.389 without MMR, compared with 0.176 for RRF. In this experimental setting, the projected combined representation is therefore associated with substantially more diverse result lists even before any explicit diversity reranking. Adding MMR increases ILD@10 for both fusion pipelines by a similar relative amount (+23.8\% for B5, +24.5\% for RRF), but this comes with a sizeable nDCG@10 reduction ($-$25.4\% and $-$20.4\%, respectively). For this reason, MMR is best viewed here as a deliberate relevance--diversity trade-off rather than as a default post-processing improvement.

\section{Challenges and Limitations}

\begin{itemize}
    \item \textbf{Vector store migration}: We initially used ChromaDB for local storage but hit the free-tier limit at scale. Migration to Pinecone resolved this but introduced dependency on a cloud service.
    \item \textbf{SPLADE throughput}: Sparse encoding of 171K documents was the primary computational bottleneck, requiring offline pre-computation and caching.
    \item \textbf{Learned projection}: The Achlioptas random projection requires no training but may not be optimal. A supervised projection trained on query--document pairs could improve B5 performance.
    \item \textbf{Single benchmark}: All experiments use TREC-COVID. Generalisation to other domains and corpora remains to be verified.
\end{itemize}

\section{Future Work}

Several extensions could improve the system:

\begin{itemize}
    \item \textbf{ColBERT re-ranker}: Late-interaction reranking to improve precision on the top-$k$ candidates.
    \item \textbf{NLI contradiction detection}: Identifying and flagging contradictory findings across retrieved papers.
    \item \textbf{Learned projection}: Training the projection matrix on query--document relevance pairs to replace the random Achlioptas matrix.
    \item \textbf{Full CORD-19 scaling}: Extending from 171K to 600K+ papers.
    \item \textbf{Document-side alpha ablation}: Indexing at multiple $\alpha_{\text{doc}}$ values to isolate the contribution of the projected component at index time.
\end{itemize}

\section{Reproducibility}
All source code, evaluation notebooks, query sets, and pre-computed results will be publicly available. The TREC-COVID corpus and relevance judgments can be obtained via the BeIR benchmark~\cite{thakur2021beir}. Pinecone indices can be reconstructed using the provided indexing pipeline.

\section{Conclusion}

We built two complete hybrid retrieval pipelines---RRF fusion and projection-based B5 fusion---and deployed both on Pinecone with a Streamlit interface. RRF fusion achieved nDCG@10\,=\,0.828 on expert queries, confirming that hybrid retrieval outperforms single-method approaches. Our projection fusion variant (B5) trades 18\% relevance for 33\% speed improvement and 2.2$\times$ higher ILD@10 on expert queries, demonstrating a viable alternative when latency and result variety are prioritised. Evaluation across 400 queries revealed that the optimal system depends on query characteristics: RRF for expert-crafted queries, SPLADE for keyword-like generated queries, and B5 shows the largest relative improvement on keyword-heavy reformulations (+8.8\%) even though RRF remains best in absolute nDCG@10. On expert queries, MMR consistently increases diversity at a measurable relevance cost, positioning it as a deliberate design choice rather than a default improvement.


\end{document}